\documentclass[useAMS,usenatbib,graphicx]{mn2e}
\usepackage{graphicx}
\DeclareGraphicsExtensions{.ps,.eps}

\def\gtsim{\raisebox{-.5ex}{$\;\stackrel{>}{\sim}\;$}}

\title[\textit{The dark burst GRB~060108.}]
  {Anatomy of A Dark Burst - The Afterglow of GRB~060108}
\author[Oates et al.]
  {S.R. Oates,$^1$\footnote{Email:sro@mssl.ucl.ac.uk.}
  C.G. Mundell,$^2$
  S. Piranomonte,$^3$
  K.L.Page,$^4$
  M. De Pasquale,$^1$
  \newauthor
  A. Monfardini,$^{2,5}$
  A. Melandri,$^2$
S. Zane,$^1$
  C. Guidorzi,$^2$
  D. Malesani,$^6$
  \newauthor
  A. Gomboc,$^{2,7}$
   N. Bannister,$^4$
A.J. Blustin,$^1$
  M. Capalbi.$^{8}$
  D. Carter,$^2$
  \newauthor
 P. D'Avanzo,$^{9,10}$
  S. Kobayashi,$^2$
 H.A. Krimm,$^{11,12}$
   P.T. O'Brien,$^4$
  \newauthor
M.J. Page,$^1$
  R.J. Smith,$^2$
 I.A. Steele,$^2$
 N. Tanvir.$^{13}$
\\
  $^1$Mullard Space Science Laboratory, University College of London,
Holmbury St Mary, Dorking, Surrey, RH5 6NT, UK \\
$^2$Astrophysics Research Institute, Liverpool John Moores University,
Twelve Quays House, Egerton Wharf, \\ Birkenhead, CH41 1LD, UK\\
  $^3$INAF, Osservatorio Astronomico di Roma, via di
Frascati 33, I-00040 Monteporzio Catone (Roma), Italy\\
$^4$Department of Physics and Astronomy, University of Leicester,
University Road, Leicester LE1 7RH, UK\\
$^{5}$ITC-IRST and INFN, Trento, via Sommarive 18, I-38050 Povo, Italy\\
  $^6$International School for Advanced Studies (SISSA-ISAS), via Beirut
2-4, I-34014 Trieste, Italy\\
 $^{7}$Department of Physics, University of Ljubljana, Jadranska 19, 6100
Ljubljana, Slovenia\\
  $^{8}$ASI Science Data Center, via G. Galilei 5, I-00044 Frascati
(Roma), Italy \\
$^9$Dipartimento di Fisica e Matematica, University
of Insubria, via Valleggio 11, I-22100 Como, Italy \\
$^{10}$INAF,
Osservatorio Astronomico di Brera, via E. Bianchi 46, I-23807 Merate (Lc),
Italy \\
$^{11}$NASA Goddard Space Flight Center, Greenbelt, MD 20771, USA\\
$^{12}$Universities Space Research Association, 10211 Wincopin Circle, 
Suite 500,Columbia, MD, 21044-3432, USA.\\
$^{13}$Centre for Astrophysics Research, University of Hertfordshire, 
College Lane, Hatfield AL109AB, UK.\\
}

\begin{document}

\date{Accepted for publication in MNRAS}

\maketitle


\begin{abstract}
We present a multiwavelength study of GRB~060108 - the 100th Gamma
Ray Burst discovered by \textit{Swift}. The X-ray flux 
and light curve (3-segments plus a flare) detected with the XRT 
are typical of \textit{Swift} long bursts. 

We report the discovery of
a faint optical afterglow detected in deep $BVRi'$ band imaging
obtained with the Faulkes Telescope North (FTN) beginning 2.75
minutes after the burst. 
The afterglow is below the
detection limit of the UVOT within 100s of the burst, while 
is evident in $K$-band images taken
with the United Kingdom Infrared Telescope (UKIRT) 
45 minutes after the burst. 
The optical light curve is sparsely sampled. Observations 
taken in the R and i$'$ bands 
can either 
be fit with a single power law decay in flux, F(t)$
\propto~t^{-\alpha}$ where $\alpha=0.43\pm0.08$, or a 2-segment 
light curve with an initial steep decay
$\alpha_1$$<$0.88$\pm$0.2, flattening to a slope
$\alpha_2$$\sim$0.31$\pm$0.12. A marginal
evidence for rebrightening is seen in the i$'$ band.

Deep $R$-band imaging obtained $\sim 12$ days post burst with the VLT
reveals a faint, extended object ($R \sim 23.5$ mag) at the location of
the afterglow. Although the brightness is compatible with the
extrapolation of the slow decay with index $\alpha_2$, significant flux is
likely due to a host galaxy. This implies that the optical light curve had
a break before 12 days, akin to what observed in the X-rays.

We derive the maximum photometric redshift $z<3.2$ for GRB~060108.
We find that the Spectral Energy Distribution at 1000~s after the
burst, from the optical to the X-ray range, is 
best fit by a simple power law,
F$_{\nu}\,\propto\,\nu^{-\beta}$, with $\beta_{OX}\,=\,0.54$ and a small 
amount of extinction. The
optical to X-ray spectral index ($\beta_{OX}$) confirm
GRB~060108 to be one of the optically darkest bursts detected. 
Our observations rule out a high
redshift as the reason for the optical faintness of GRB~060108. We 
conclude that a more likely explanation is a combination of an intrinsic 
optical faintness of the burst, an hard optical to X-ray spectrum 
and a moderate amount of extinction
in the host galaxy.

 \end{abstract}

\begin{keywords}
gamma-rays:bursts  
\end{keywords}

\section{Introduction}

Gamma ray bursts (GRBs) are brief, intense and totally unpredictable
flashes of gamma rays on the sky that are thought to be produced
during the core collapse of massive stars (long-duration bursts) or
the merger of two compact objects such as two neutron stars or a
neutron star and stellar-mass black hole. Until the
recent launch of the {\it Swift} satellite in November 2004, it was
notoriously difficult to observe GRBs at other wavelengths within
seconds or minutes after the burst. Nevertheless, the successful
identification with BeppoSAX of bright, long-lived X-ray afterglow
emission for long bursts \citep{cos97} and that of the 
corresponding optical and infrared counterparts
\citep{vp97}, established GRBs as cosmological, and therefore the
most instantaneously luminous objects in the Universe. Similar
breakthroughs for short bursts have recently occurred, showing them
also to be extragalactic, but less luminous and less distant than
long bursts \citep{geh05,vil05,fox05,cov06,bart05}.

With the availability of \textit{Swift}'s
promptly-disseminated arcsec localizations and the 
on-board rapid-slew X-ray and
ultraviolet/optical telescopes (XRT, UVOT; \citealt{geh04}),
multi-wavelength monitoring of GRBs from the earliest possible times
is now being performed for a significant number of bursts.
Additionally, large aperture ground-based robotic telescopes such as
the 2-m Liverpool \citep{steele04} and Faulkes telescopes respond
rapidly to GRB alerts and begin automatically imaging the target
field within minutes of receipt of an alert, providing early deep
upper limits or multi-colour follow-ups of optical counterparts as
faint as R\,$\sim$\,18\,-\,22\,mag (e.g. Guidorzi et al. 2005a,
2006b; Monfardini et al. 2006b).

Despite increasingly rapid responses that provide sensitive limits
within minutes of the burst, the absence of long-wavelength emission
afterglows for a significant number of GRBs (so-called ``dark
bursts'') remains a key unsolved problem. In the pre-\textit{Swift}
era, as many as 50\% of BeppoSax bursts were lacking an optical detection
\citep{dp03,lcg02}. The discovery rate of optical afterglows was even  
higher for HETE2 than for SAX (\citealt{lamb04}), and it was expected 
to increase significantly in the \textit{Swift} era of rapid followup. 
Instead,
a substantial fraction of \textit{Swift} bursts remain 
undetected in the optical band \citep{rom05}. 
Possible scenarios to explain the ``observed'' optical darkness of
these bursts, apart from fast-fading transients lacking sufficiently
deep, early-time observations \citep{groot98}, include
intrinsically-faint optical afterglows \citep{fyn01, lcg02, dp03,
ber05}, highly obscured afterglows whose optical light is absorbed
by the circumburst or interstellar material \citep{lcg02,dp06}, high
redshifts \citep{fru99,lr00,bl02,ber02,tag05}, and radiative
suppression in a sub-class of bursts with unusually high
$\gamma$-ray efficiency producing intrinsically low X-ray and
optical fluxes \citep{rom05,ped06}.

 Here we present a multi-wavelength X-ray, optical and infrared study
of the optically-faint GRB~060108. The Burst Alert Telescope (BAT,
\citealt{bart05}) was triggered by this GRB - {\it Swift}'s 100th burst
- at 14:39:11.76 UT on January $\rm 8{^t}{^h} $ 2006. The
$\gamma$-ray light curve has a single peaked structure with a FRED
time profile \citep{oat06}, a duration
T$_{90}$\,=\,14.4\,$\pm$\,1.6\,s (in the 15\,$-$\,350\,keV band),
and a 15\,$-$\,150\,keV fluence $S_\gamma = (3.7 \pm 0.4) \times
10^{-7} {\rm erg\,cm}^{-2}$. Here and in the following errors are at 90\% 
confidence level, unless specified otherwise. 
 The \textit{Swift} X-ray and ultraviolet/optical telescopes (XRT/UVOT) began
observing at 91\,s and 76\,s after the BAT trigger respectively,
followed shortly after by the Faulkes Telescope North (FTN) at 2.75
minutes post trigger. Infrared observations with the United Kingdom
Infrared Telescope (UKIRT) were acquired during the optical imaging
period beginning at 45 min postburst. The Very Large Telescope (VLT)
was used to obtain further near infrared imaging at 16.1 hours and
2.7 days. Following an initial estimate of the position of the X-ray
afterglow \citep{pag06}, a revised location was derived
\citep{bb06}; the location of the optical \citep{mon06a} and
infrared counterparts \citep{dav06,lev06} was found to be consistent
with this revised XRT position. Deep $R$-band imaging and spectroscopy
were subsequently performed with the VLT at 12.7 days and $\sim$\,21 days
respectively.

\section{Observations and Analysis}
\label{sec:observations}

\subsection{\textit{Swift} Gamma-Ray Observations}
\label{sec:batobs}

At 14:39:11.76\,UT, on $\rm 8^{th}$ January 2006, the \textit{Swift}
BAT triggered and located GRB~060108 (BAT trigger 176453)
\citep{oat06}. Unless otherwise specified, times in this section are
referenced to the BAT trigger time, hereafter designated $T_0$. The
burst was detected in the fully coded part of the BAT field of view,
meaning that all BAT detectors were illuminated by the source. The
spacecraft began to slew to the source location 11.2~s after the
trigger and was settled at the source location at $T_0$+73.20\,s.

The BAT data for GRB~060108 between $T_0$-300\,s and $T_0$+300\,s
were collected in event mode with 100\,$\mu$s time resolution and
6\,keV energy resolution. The data were processed using standard
\textit{Swift}-BAT analysis tools and the spectra were fit using
XSPEC\,11.3. Before processing, the BAT event data were corrected
for mask tagged weighting using the standard BAT
tool, \textit{batmaskwtevt}. This procedure scales each count in each
detector by an amount proportional to the portion of the detector
exposed to the source through the empty spaces in the BAT coded
mask. Mask weighting effectively removes the background and all BAT
light curves in this paper have been background subtracted by this
method.

\subsection{\textit{Swift} X-ray Observations}
\label{sec:xrtobs}

The X-ray Telescope (XRT, \citealt{bur05}) position
refined analysis located this burst at $\alpha$(J2000)\,=
\,09$^{\rm h}$48$^{\rm m}$01\fs6, 
$\delta$(J2000)\,=\,31\degr55\arcmin04\farcs6 
\citep{pag06}.
XRT began observations approximately 90\,s after the burst trigger.
The analysis of the XRT data was performed using the XRT pipeline
software version 2.3. The initial exposure was taken in Image Mode
and no source was detected. The XRT then observed in Windowed Timing
(WT) mode for $\sim$\,10\,s before switching to Photon Counting (PC)
mode for a further 10\,ks; a fading uncatalogued source was
detected. {\it Swift} observed the field until 10$^6$\,s after the
burst trigger.

The source and background counts used to create the light curve and the spectra
were obtained from the cleaned event files using extraction regions.
The analysis of the WT mode data provided only limited results as
the exposure was of such a short duration.

The first 1000\,s of PC data were found to be piled up. The size of
the region that was affected by the pile-up was determined by
comparing the measured point spread function (PSF) with a model XRT
PSF. The affected radius was found to be 1.6 pixels (4\arcsec) in
radius, therefore an annular region with inner radius 4\arcsec and
outer radius of 30 pixels (70.8\arcsec) has been used for the
extraction. The rest of the PC data were extracted using a circular
region of radius 30 pixels (70.8\arcsec) and a background region of
radius 80 pixels (189\arcsec). These regions were used to extract
data from all orbits of PC data.

When extracting the PC mode spectrum, only events with grade
0\,-\,12 were used, and for the WT mode, events were used of grade
0\,-\,2. The energy response of the detector was taken into account
when fitting the spectrum. The response matrices (RMs) were taken
from the {\it Swift} calibration database, CALDB 20060426. These were the 
most recent calibration products
available and were released in April 2006. The effective area files
were created with {\it xrtmkarf} and required an image file created
using {\it expomap}. This code maps the detector plane so any bad
pixels or hot columns are accounted for in the effective area file.
A correction factor calculated with {\it xrtmkarf} was applied to
the piled up section of the XRT light curve.

During the observations, the burst's location near to the Sun caused
the XRT to warm up, leading to mode switching of the instrument.
This only affected data in the second orbit and onwards. Thus the PC
light curves, where appropriate, were also corrected for low
fractional exposure.

\subsection{Optical and Near Infrared Imaging}
\subsubsection{\textit{Swift} Ultraviolet Optical Telescope (UVOT)}
\label{sec:uvotobs}

The UVOT began observations $75$\,s after the BAT trigger. The first
$\sim$\,11\,s exposure, taken while the spacecraft was settling on the
target, was taken through the $V$ filter in photon-counting (``Event'')
mode. Once the pointing had stabilized, a 200\,s $V$-band finding-chart
exposure in Image/Event mode was taken. After this, the UVOT cycled
through the colour filters. The later exposures were in image mode only.
Counts were extracted from the images using a 3\arcsec\ radius aperture at
the position of the optical afterglow given in this paper. The count rates
were aperture-corrected to ensure compatibility with the photometry
calibration by \cite{poo05}.
We then derived the 3$\sigma$ upper limits in
count units, which were converted to magnitudes using the zero-points
available in the \textit{Swift} CALDB, and corrected for Galactic
extinction using $E(B-V) = 0.018$. Results are given in Table~\ref{uvot}.

\begin{table}
\caption{{\it Swift} UVOT observations of GRB~060108, corrected for
Galactic extinction. $E(B-V) = 0.018$.} \label{uvot}
\begin{tabular}{|l|c|c|c|c|}
\hline
Filter & Time Range & Exposure & 3$\sigma$ Upper Limits \\
& (s) & (s) & (mag)\\
\hline
$V$ & 75-86   & 11    & 17.9 \\
$V$ & 90-290  & 200   & 19.6 \\
$V$ & 76-1096 & 357 & 19.9\\
$V$ & 16835-17735 & 900  & 20.8 \\
$B$ & 457-886 & 100 & 20.1\\
$B$ & 10983-40489  & 1588  & 21.8 \\
$U$ & 403-832 & 100 & 20.2\\
$UVW1$ & 349-5624 & 521 & 20.9\\
$UVM2$ & 295-5195 & 999 & 21.7\\
$UVW2$ & 566-995 & 100 & 20.2\\
\hline
\end{tabular}
\end{table}

\subsubsection{Faulkes Telescope North (FTN)}
\label{sec:ftnobs}

\begin{figure}
\includegraphics[width=85mm]{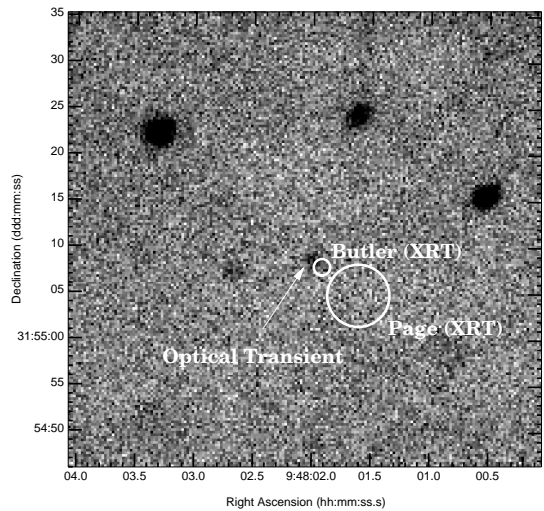}
\caption{Faulkes Telescope North i$'$-band
image of the optical counterpart to  GRB~060108 (indicated with an arrow). The
original \citep{pag06} and revised \citep{bb06} XRT
error circles are also shown. }
 \label{fig:ftnimage}
\end{figure}

The Faulkes Telescope North, located on Maui, Hawaii, responded
robotically to the \textit{Swift} alert and began observing 2.75~min
after the GRB trigger. The initial detection mode images (three 10~s
$R$-band exposures) were processed automatically by the Liverpool GRB
pipeline, LT-TRAP \citep{g06a}. No new source has been identified in
the XRT error box \citep{pag06} brighter than $R \sim$\,19.5~mag
($\Delta$t\,=\,3.1 min) and $R \sim$\,20.2~mag ($\Delta$t\,=\,11.7
min) in comparison with USNO-B1.0, 2MASS and GSC 2.3 catalogues
\citep{g06b}. The telescope continued in multicolour imaging mode,
obtaining the exposures in $BVRi'$ bands.

\begin{table*}
\begin{minipage}{126mm}
  \caption{Optical detections and
lower limits. Errors are at 90$\%$ confidence level.}
  \label{tab:resoptir}
  \begin{tabular}{@{}lccccccc}
   \hline
Mean Date&Filter &Telescope     & Exposure & Seeing&Magnitude& Time Range
&
$\Delta$t \\
 (UT)       &     &       & Time (s)&($\prime\prime$)&   & (s) & (s) \\
\hline
Jan 8.613776 &Bessell-B & FTN   &   10.0  & 1.12     & $>$19.90 &
269$-$279&274  \\
Jan 8.624165 &Bessell-B & FTN   &   390.0  & 1.54    & 22.31$\pm$0.26 &
450$-$1552& 1172\\
Jan 8.643794 &Bessell-B & FTN   &   540.0  &1.74     & 22.72$\pm$0.35 &
2198$-$3310& 2868\\
Jan 8.614304 &Bessell-V & FTN   &   10.0   & 1.29    &
21.7$\pm$1.2$^{\dagger}$& 315$-$325&320 \\
Jan 8.612498 &Bessell-R & FTN   &    10.0  & 1.09    & $>$20.20       &
159$-$169& 164\\
Jan 8.612996 &Bessell-R & FTN   &    10.0  & 1.09    & $>$20.20       &
202$-$212&207 \\
Jan 8.616757 &Bessell-R & FTN   &    30.0  & 1.18    & 21.23$\pm$0.44 &
517$-$547& 532 \\
Jan 8.620774 &Bessell-R & FTN   &   240.0  &1.47     & 21.84$\pm$0.28 &
164$-$1186& 879 \\
Jan 8.633794 &Bessell-R & FTN   &   300.0  & 1.55    & 21.88$\pm$0.31  &
1767$-$2359 &2004\\
Jan 8.649744 &Bessell-R & FTN   &   420.0  &1.74     & 21.98$\pm$0.31 &
2944$-$3711& 3382\\
Jan 8.614940 &SDSS-i$'$ &FTN    &    10.0     & 1.12  &
20.9$\pm$0.7$^{\dagger}$ & 370$-$380& 375\\
Jan 8.619292 &SDSS-i$'$ &FTN    &   100.0     & 1.34  & 21.69$\pm$0.38 &
375$-$891& 751   \\
Jan 8.630704 &SDSS-i$'$ &FTN    &   300.0     & 1.51  & 21.45$\pm$0.16 &
1352$-$1993& 1737\\
Jan 8.644338 &SDSS-i$'$ &FTN    &   300.0     & 1.56  & 21.88$\pm$0.33 &
2464$-$3265 & 2915\\
   \hline
  \end{tabular}
  \medskip\\
$^{\dagger}$Marginal detection with low statistical significance.
Errors are derived by adding statistical and systematic errors in
quadrature. Magnitudes are uncorrected for Galactic extinction.
3$\sigma$ upper limits are quoted for non-detections. Column 7 lists
the Time range covered by the exposures and column 8  lists the
corresponding exposure-weighted mean time  ($\Delta$t) since the
burst for co-added images calculated as
$\Delta t = (\sum_i t_i \Delta t_i) / (\sum_i \Delta t_i)$.
\end{minipage}
\end{table*}

The robotic followup ended after one hour.
Subsequent post-processing of the $i'$ band images revealed an
uncatalogued source $\sim$3\arcsec\ NE of the edge of the XRT error
circle (see Figure \ref{fig:ftnimage}). The $i'$-band brightness of
this source in the FTN image relative to the neighbouring object was
greater than in the pre-burst SDSS $i'$-band image, suggesting it as
the candidate afterglow. Subsequent analysis of the XRT data by
\citet{bb06} confirmed an XRT offset with a revised position for the
afterglow $\alpha$(J2000)\,=\,09$^{\rm h}$48$^{\rm m}$01\fs92,
$\delta$(J2000)\,=\,31\degr55\arcmin07\farcs8 
with 0\farcs9 uncertainty (90\% containment). As can be seen in Figure
\ref{fig:ftnimage}, the location of the optical afterglow at
$\alpha$(J2000)\,=\,09$^{\rm h}$48$^{\rm m}$01\fs98,
$\delta$(J2000)\,=\,31\degr55\arcmin08\farcs6 with $\sim$\,0\farcs5
uncertainty (90\% containment, \citealt{mon06b}), is consistent with the 
revised XRT position. A
similar location is derived for the NIR counterpart identified by
\citet{dav06} and \citet{lev06}. Careful analysis of the optical
data subsequently confirmed that the source was fading and further
imaging with the FTN and other telescopes was manually triggered to
follow the later-time evolution. Unfortunately moon and weather
constraints prevented deep detections or limits beyond $R \sim$\,23
mag.

Table \ref{tab:resoptir} summarizes the results of the optical and
NIR imaging; as no FTN observations of standard stars were
available, magnitudes in $B$, $V$, $R$, and $i'$ were calibrated as
follows.
Observations in $i'$ band were calibrated directly using the SDSS DR4
photometry\footnote{www.sdss.org/dr4/}, while $B$, $V$, and $R$ band data
were calibrated assuming the pre-burst SDSS $g'r'i'$ photometry of
\cite{cool06a,cool06b} combined with observed filter transformations
given by \citet{sm02} in their Table 7. Finally, the data were
corrected for the Galactic extinction, which is low towards GRB~060108:
A$_B$~=~0.09~mag,
A$_V$~=~0.07~mag, A$_R$~=~0.05~mag, A$_i$~=~0.04~mag
\citep{sch98,car89}.
 Magnitudes were converted into
flux densities F ($\mu$Jy) following \citet{bes79} and \citet{f96}
for comparison with X-ray flux densities.

\subsubsection{Very Large Telescope (VLT)}
\label{sec:vltobs}

$J$- and $K_{\rm s}$-band observations of the field of GRB~060108
were performed by using the ISAAC camera at two different epochs,
and in the $R$ band by using the FORS\,1 instrument at one epoch
(see Table \ref{tab:vltphot}). Both instruments are mounted on the
ESO-VLT located at Cerro Paranal (Chile). All nights were
photometric, with a seeing in the range 0.8\arcsec--1.4\arcsec.
Image reduction was performed following the standard procedures.
Astrometric solutions were computed using the USNO
catalog\footnote{http://www.nofs.navy.mil/data/fchpix/}.
Aperture photometry was performed using SExtractor (\citealt{Be96}). In
the optical, the flux calibration was performed observing Landolt
standard stars. In the NIR, the 2MASS catalogue (\citealt{2mass}) was 
adopted as a reference, even
if only a handful of usable stars were present in the ISAAC field.
Late-time low-resolution spectra were acquired with the VLT in an attempt
to  measure the host galaxy redshift. However, the object was quite faint
and spectra have a very low signal-to-noise ratio, with no clear emission
or  absorption lines. We will not discuss further these spectra.

In Table~\ref{tab:vltphot} we present the observing log and list the
results of our spectrophotometric analysis for each epoch.

\begin{table*}
\begin{minipage}{126mm}
\caption{GRB~060108: VLT observation log. Errors are at 90$\%$
confidence level, while 3$\sigma$ upper limits have been used for 
the non-detection.
\label{tab:vltphot}}
\begin{tabular}{lcccccc}
\hline
Mean UT   &Time since
& Exp                   & Seeing   & Instrument  & Filter/    & Magnitude
             \\
            &burst (d)    &(s)
&(\arcsec) &             & grism      &
             \\ \hline
Jan 9.307 &  00.696     & 10$\times$90         & 1.1      & ISAAC
& $J$ & $> 22.3$
             \\
Jan 11.283 &  02.672     & 60$\times$60          & 0.8      & ISAAC
&$K_{\rm s}$ & 20.49$\pm$
0.15     \\
Jan 21.286 &  12.676     & 10$\times$120         & 1.4      & FORS1
& $R$        & 23.46$\pm$0.15     \\ \hline Jan 29.237 & 20.626
& 3$\times$2160        & 0.8       & FORS2      & 300V &
---
          \\
Jan 30.176 &  21.565
&  3$\times$2160        & 0.8       & FORS2      & 300V       & ---
          \\ \hline
\end{tabular}
\medskip\\
\end{minipage}
\end{table*}

\section{Results}
\subsection{GRB~060108 and Its Afterglow}

The afterglow of GRB~060108 is detected at X-ray \citep{pag06},
optical \citep{mon06a} and near infrared wavelengths \citep{lev06}.
The X-ray light curve shows the canonical 3-segment decay similar to
many gamma ray bursts detected with \textit{Swift}, with a small
flare 300\,s after the burst. Optically, the burst is not detected
at ultraviolet and optical wavelengths with the UVOT in the first
two minutes, but it is detected in deeper $B$, $V$, $R$ and $i'$ band FTN
images as early as 5.3~min after the burst. The optical afterglow of
GRB~060108 is one of the faintest yet discovered. 

\begin{figure}
  \includegraphics[angle=0,scale=0.45]{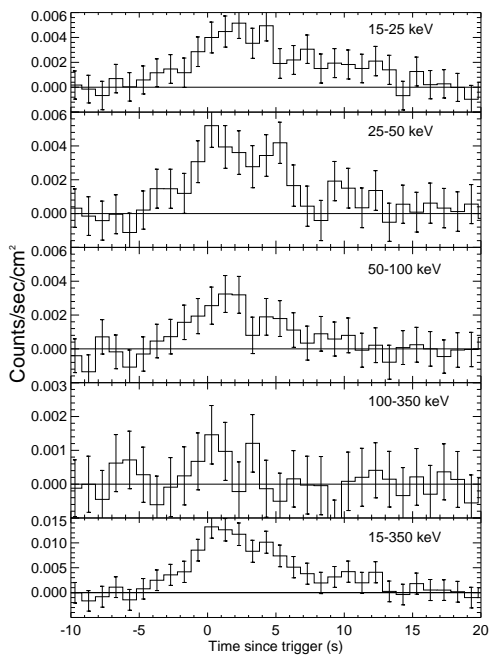}
  \caption{BAT light curve of GRB~060108 in different energy bands:
15\,-\,25, 25\,-\,50, 50\,-\,100, 100\,-\,350\,keV, and total range  
15\,-\,350\,keV.}
\label{fig:b1curve}
\end{figure}

\subsection{{\it Swift } Gamma-Ray Light curve and Spectra}
\label{sec:resultsgammaray}

The BAT light curve (see Fig.~\ref{fig:b1curve})
shows a simple triangular profile in all energy bands, with a rise
from background starting at $T_0$--4.0\,s, a peak at $T_0$, and a
return to background levels by ~$T_0$+13.2\,s. A small
secondary peak is seen at around $T_0$+10\,s, but then no detectable emission has been
registered after $T_0$+13.2\,s nor at the times of the
flare seen by the XRT. The peak count rate measured by BAT is
2000 counts\,s$^{-1}$ at $T_0$, and the duration is $T_{90}=14.4 \pm 
1.6$~s
(15\,-\,350\,keV band; estimated error including systematics). The 1s
peak photon flux measured from $T_0$--0.48\,s
is 0.7\,$\pm$\,0.1 ph\,cm$^{-2}$\,s$^{-1}$ (in the 15\,-\,150\,keV band).

For the spectral analysis, we have analyzed the data between 15 and
150\,keV, where BAT has a better response. We find that the emission
from GRB~060108 shows the hard to soft spectral evolution commonly
observed in many GRBs and that, for all time periods, the best fit
to the BAT data is a simple power law. The spectrum of the first
peak ($T_0$-4.02\,s to $T_0$+9\,s) is best fitted with a photon
index of 1.9\,$\pm$\,0.17
 ($\chi^2$/d.o.f.\,=54/90), while that of the smaller second peak
($T_0$+9\,s to $T_0$+13.2\,s) has a photon index
2.55\,$\pm$ \,0.61 ($\chi^2$/d.o.f. = 64/90). With the given spectral
parameters, this burst appears to share some similarity with the X-ray
flashes (XRFs). Although \textit{Swift} is
not well suited to identify XRFs, since it lacks  coverage at low energy
(i.e. in the the ``classical'' HETE band 2-30 keV), we notice that,
if the spectrum is extrapolated down to this band, then the soft photon
index would imply an XRF or at least an X-ray rich event.

\subsection{{\it Swift } X-ray Light curve and Spectra}
\label{sec:resultsgx}

\begin{figure}
  \includegraphics[angle=-90,scale=0.3]{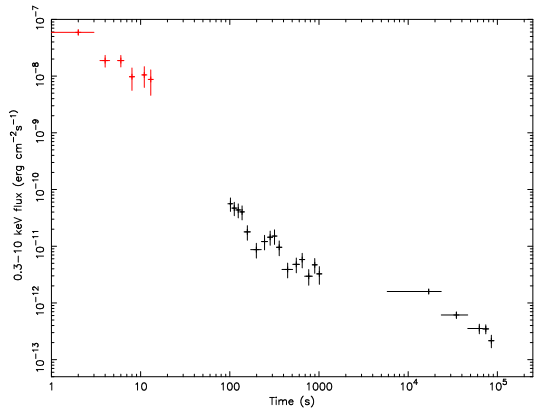}
  \caption{BAT and XRT combined unabsorbed flux light curve of
GRB~060108.}
\label{fig:batxrtcurve}
\end{figure}

By following \cite{obr06}, the BAT light curve can be extrapolated
at lower energies and joined to the XRT light curve in a single
plot. Although some consideration is needed (as the energy bands of
the two instruments do not overlap), this is a powerful method which
allows us to visualize the decay rate of the event through a longer
time interval, commencing at the burst trigger. The combined BAT and
XRT light curve resulting from this process is shown in Figure
\ref{fig:batxrtcurve}. For the extrapolation, we used a BAT photon 
index of $1.9$ (computed in the 15-150~keV band).

The light curve has a complex shape, showing a steep decay, followed by a
flatter phase which eventually steepens again. A visual inspection shows a 
possible flare, superimposed
between the two segments.
We modeled the X-ray light curve with both a double broken power-law
and a double broken power-law with the inclusion of a Gaussian, to
represent the flare. In the first case we find a
$\chi^2$/d.o.f.\,=\,26/23, while by using the second model gives
$\chi^2$/d.o.f\,=\,11/20.  The probability that this improvement is
due to chance, calculated using the F-test, is  0.05$\%$. 
In order to check for overfitting effects, we repeated the fitting by 
adopting different choices for the temporal binning. We find that, in all 
cases, the light curve is better fitted by including a Gaussian, although in 
some cases both fits  
(with and without Gaussian) have reduced $\chi^2$ values larger than unit, 
and in turn the decrease in $\chi^2$ is less pronunced. The 
probability that the improvement is
due to chance, calculated using the F-test, varies between  0.019 and 
0.035 in the different fits, which means that the fit including the 
Gaussian can be used with 
a confidence of $\gtsim 2\sigma$. The 
best fit parameters appear to be always consistent with each others, 
and are given in
Table~\ref{tab:tempindices}; here and in the following we assume the
convention $F \propto t^{-\alpha}\nu^{-\beta}$. 

\begin{table}
 \caption{Best fit parameters for a fit of the XRT light curve of
 GRB~060108 with a double broken power law model plus a Gaussian.
The epoch of the flare, $t_{\rm flare}$, and its width $\sigma_G$ have 
been reported. From the Gaussian component, the Full Width Half Maximum is 
related to $\sigma_G$ by FWHM$=2.35 \sigma_G$.  
Errors are at 90$\%$ confidence level.}
 \label{tab:tempindices}
\begin{tabular}{@{}lc}
\hline
Parameter & \\
\hline
$\alpha_1$  & $2.78 \pm 0.43$ \\
$\alpha_2$  & $0.30 \pm 0.07$\\
$\alpha_3$ & $1.05 \pm 0.10$ \\
$t_{\rm break,1}$  & $250 \pm 30$~s\\
$t_{\rm break,2}$  & $(11.5 \pm 0.4) \times 10^3$~s \\
$t_{\rm flare}$  &$300 \pm 10$~s \\
$\sigma_G$   & $32.3 \pm 10.0$~s\\
\hline
\end{tabular}
\medskip\\
\end{table}

\begin{table*}
 \caption{X-ray spectral indices for GRB~060108.
Errors are at 90$\%$ confidence level. When possible, we report the
90\% upper limit on the excess in column density, over the Galactic
value of $N_{H}^{Gal}=1.7\times10^{20}$cm$^{-2}$ (\citealt{dic90}).
The
energy index of the first and second segment have been computed
twice: with ($\beta_{i}$) and without ($\beta_{i}'$) the inclusion
of the X-ray flare. In the second case Cash statistics has been
used.}
 \label{tab:xindices}
\begin{tabular}{@{}lcccc}
\hline Segment & Interval & $\beta_i$ & $N_H$ excess & $\beta_{i}'$ \\
& & & cm$^{-2}$ &  \\
\hline
1 & $t < t_{\rm break,1}$  & 1.31 $\pm$ 0.44 & &  $1.47^{+0.43}_{-0.41}$
\\
Flare&    &   -        & & 1.7 $\pm$ 0.4 \\
 2 &$t_{\rm break,1} < t < t_{\rm break,2}$
& 0.68 $\pm$ 0.23 & $< 1.4 \times 10^{21}$ & 0.89 $\pm$ 0.3 \\
3&$t > t_{\rm break,2}$  & 0.71 $\pm$ 0.15 & $< 1.7 \times 10^{21}$ & \\
\hline
\end{tabular}
\medskip\\
\end{table*}

Our analysis shows that there are three distinct segments in the
X-ray light curve. The X-ray afterglow starts with a phase of steep
decay (first segment, slope $\sim 2.8$), followed by a shallow
section (second segment, slope $\sim 0.3$), and finally enters a
``normal afterglow'' decay phase, with a slope $\sim 1$ (third
segment). The two break times which separate these three segments
are $t_{\rm break,1} \sim 250$~s and $t_{\rm break,2} \sim 11.5$~ks.
This behaviour is typical of X-ray afterglow lightcurves as shown by 
\cite{no05}.

In order to perform a spectral analysis, we fitted the spectra taken
during different time intervals by using an absorbed power-law
model. The Galactic value of the column density has been fixed at
$N_{H}^{Gal}=1.7\times10^{20}$cm$^{-2}$ (\citealt{dic90}; this
value is in agreement with that obtained from the more recent LAB
survey,
which is $1.8\times 10^{20}$cm$^{-2}$, \citealt{kalb05}
\footnote{http://www.astro.uni-bonn.de/$^\sim$webrai/english/tools\_labsurvey.php.}), and an
additional absorption component, $N_H$, has been allowed to mimic
the intrinsic excess in column density. Results are summarized in
Table~\ref{tab:xindices}.

The spectral indices before and after the first break at
250\,$\pm$\,30\,s have been first computed by including the flare at
\,300\,$\pm$\,10\,s ($\beta_i$ in the table). In the attempt to
assess the flare contamination on these values, we divided the
interval in three parts by isolating the time interval of the flare.
Since these new spectra contained less than ~100
background-subtracted counts, Cash statistics were used for the
fitting (a minimum of 15-20 counts per bin are required for Gaussian
$\chi^2$ statistics). A comparison of the resulting indices,
$\beta_i'$, with the $\beta_i$ previously computed, shows that the flare
does not alter
significantly the spectral indices registered during the first two
decay phases.

Because of the low statistics, when considering the spectra of
individual segments separately we are unable to constrain the value
of the excess of column density over $N_{H}^{Gal}$. We have only
been able to derive a 90\% upper limit, and only by using the
spectra of the last two time intervals.
In order to have an estimate for the photoelectic absorption, we
re-extracted the X-ray spectrum using all
data after the flare through to the end of the
observations (time interval 570~s$<t<4.63 \times 10^5$~s). Since this
spectrum has a better signal to noise, the excess
$N_H$  is actually significant at 99\% confidence level; the
best-fit spectral parameters are $\beta_i = 1.06^{+0.27}_{-0.23}$,
$N_H = 1.02^{+0.75}_{-0.66}\times 10^{21}$~cm$^{-2}$ (see
also \S\ref{dark}).

\subsection{Optical and Infrared Properties}
\label{sec:opticalIR}

\begin{figure}
\includegraphics[width=60mm,angle=270]{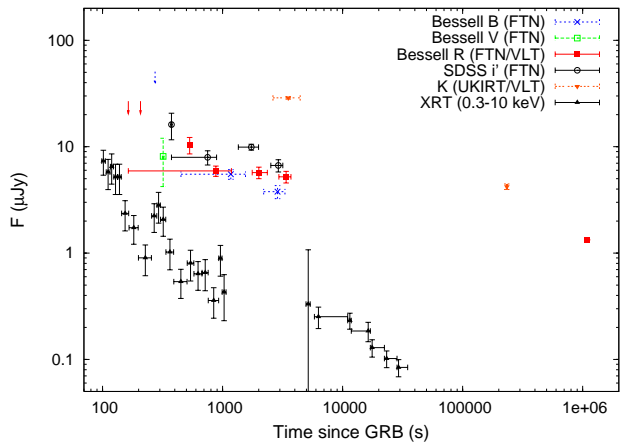}
\caption{$BVRi'$ multicolour light curves of GRB~060108 obtained by
the FTN plus $K$-band detections from UKIRT (value taken from
\citealt{lev06b}) and the VLT. Flux densities have been corrected
for Galactic extinction. The XRT light curve is also shown for
comparison.}
 \label{fig:optircurve}
\end{figure}

The optical and infrared properties of the GRB~060108 afterglow are
also particularly interesting. Figure \ref{fig:ftnimage} shows the
identification of the optical afterglow with the X-ray counterpart
in an FTN $i'$-band image, while Figure \ref{fig:optircurve} shows the
optical and infrared light curves including detections and upper
limits from optical to near infrared wavelengths, corrected for
Galactic extinction. As shown in Table \ref{tab:resoptir}, the
afterglow is visible in all $BVRi'$ bands from as early as
$\Delta$t\,$\sim$\,5 minutes, but is not detected by UVOT within the
first few minutes of the trigger or during the short exposure FTN
$R$-band and $B$-band images taken at 164\,s and 274\,s after the
trigger, respectively. Although these FTN exposures have only a
$\sim 10$\,s duration, the deep limits rule out a significant
optical flare coincident in time with that possibly seen in the X-ray light
curve. Moreover, the marginal FTN $V$-band detection (Table
\ref{tab:resoptir} and Figure \ref{fig:optircurve}) reveals that the
afterglow was already fainter than the UVOT limiting magnitudes.

\begin{figure}
\includegraphics[width=80mm]{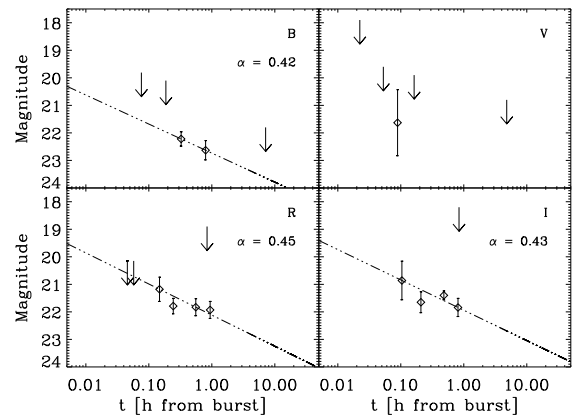}
 \caption{Best fitting power law
 curves for the afterglow as observed with the
FTN, consistent with the gradient between the 2 NIR datapoints (not shown).
Upper limits (including those registered by the UVOT) are
indicated with arrows.}
 \label{fig:curvefit}
\end{figure}

The optical light curves to 4000\,s after the burst can be fit with a
single power law decay in flux density $F(t)\,\propto\,t^{-\alpha}$ where
$\alpha\,=\,0.43\,\pm\,0.08$ (Figure \ref{fig:curvefit}), a slope that is
also consistent with the gradient between the two $K$-band detections.
Alternatively, guided by the structure in the X-ray light curve and
concentrating on the $R$ and $i'$ band light curves, a more complicated
interpretation is possible. By a direct comparison between the X-ray and
optical flux density evolution (see Figure~\ref{fig:optircurve}) we find
that, although the optical light curves are sparsely sampled (due to the
need to coadd frames), there is evidence for a 2-segment light curve. A
possible rebrightening is seen in $i'$ band, although the significance is
low and a monotonic decay is not ruled out. Using the early time $R$-band
upper limit as a constraint on the maximum possible value of $\alpha$ and
fitting the $R$-band light curve in two time portions $300~{\rm s}< t_1 <
1000~{\rm s}$ and $t_2 >800$\,s, we derive $\alpha_1\,<\,0.88\,\pm\,0.2$
and $\alpha_2\,=\,0.31\,\pm\,0.12$. The second decay
$\alpha_2$ is found to be shallow whether or not the late time VLT
$R$-band measurement at 12.68 days is included and, by assuming that a
break has not occurred in the optical light curve, this may represent a
faint detection of the afterglow. On the other hand, as suggested by the
X-ray light curve, it is possible that a break has already occurred around
$t_{\rm break,2} \sim 11.5$~ks, in which case the predicted magnitude of
the optical counterpart would be significantly fainter than $R\sim$\,23.5
mag at 12.68 days. As discussed in \S~\ref{sec:discussion}, if taken 
at face value the apparent rebrightening in the $i'$-band light curve 
(Figure
\ref{fig:optircurve}) may be interpreted as due to a further emission
mechanism (possibly a second reverse shock). In such a case, the decay at
the end of the first hour may be significantly steeper than
$\alpha_2$\,$\sim$\,0.31, similar instead to $\alpha_1$. Moreover, in the
late VLT image the object is not pointlike but clearly extended, making
likely that the emission detected at such late time is dominated by the
host galaxy (at least $\sim50\%$ of the total flux).

\subsubsection{Photometric Redshift Derivation}
\label{redshift}

\begin{figure}
\includegraphics[width=80mm,height=60mm]{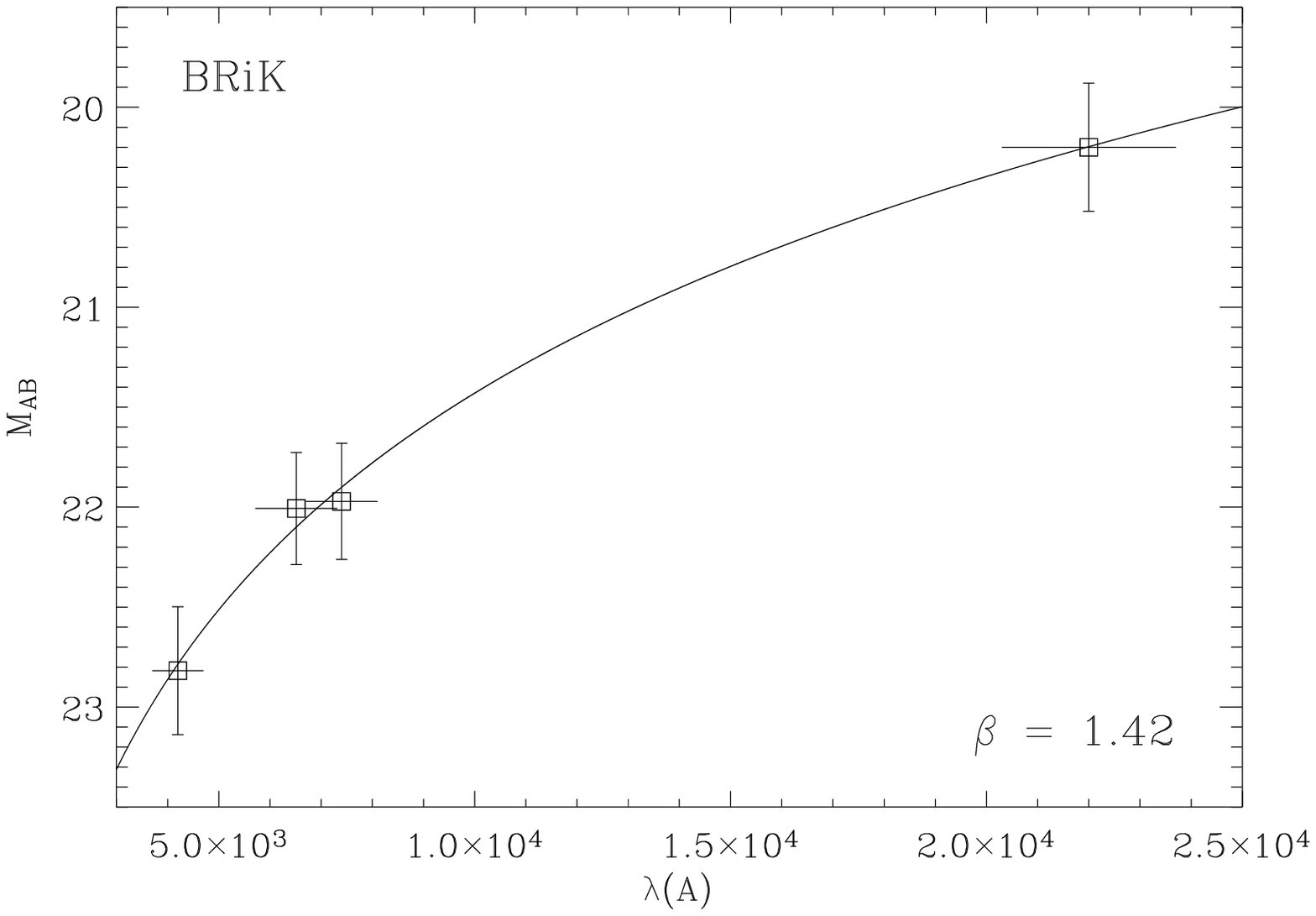}
\includegraphics[width=80mm,height=60mm]{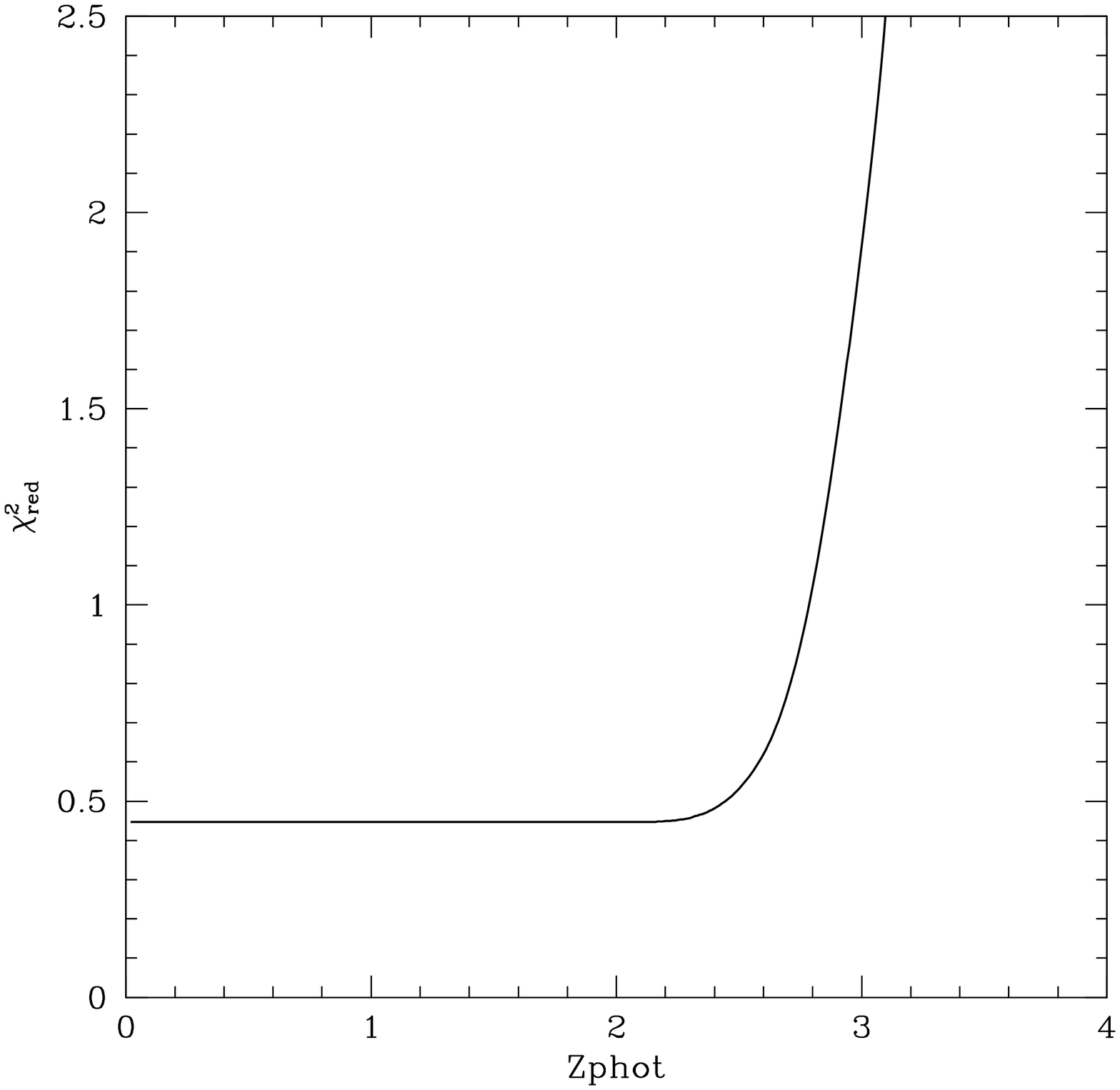}
 \caption{{\it Top panel}: the optical:NIR spectral energy distribution
for contemporaneous magnitudes in $B$, $R$, $i'$ and $K$ bands.  Observed
values
were extrapolated to 3600~s after the burst by using a decay slope
of $\alpha$\,=\,0.4, coincident with that of the optical temporal decay.
{\it Bottom panel:} the reduced $\chi^2$ distribution as a function of
photometric redshift. See text for details.}
 \label{fig:redshift}
\end{figure}

In order to estimate the photometric redshift of GRB~060108, we
adopted a $\chi^2$ minimization of the observed optical to infrared
Spectral Energy Distribution (SED) based on the technique presented
by \citet{f00}. This is a widely used and well tested technique for
redshift determination of galaxies (in which case it takes into
account for star formation history of each galaxy type and other
effects; \citealt{fs99}, \citealt{cs00}, \citealt{rr03}, \citealt{mel06}) 
which has been recently
readapted to fit simple power-law SEDs, as those typical of GRB
afterglows.

The photometric redshift was determined by combining contemporaneous
optical and NIR flux measurements derived from the early-time FTN and
UKIRT imaging at 45 minutes after the burst.
The $BRi'K$ values were extrapolated to 3600~s using
different values for the temporal decay index ranging from $\alpha$\,=0.3
to $\alpha$=0.6 (see Figure~\ref{fig:redshift}). Following the method
described in \citet{mel06}, the
magnitudes were
converted to the AB system before inputting them into the photometric
redshift determination algorithm.

Adopting a $\chi^2$ minimization of the observed spectral energy
distribution (which is assumed to
follow a power law F$_{\nu}\,\propto\,\nu^{-\beta}$) 
at 3.6\,ks after the trigger, we derive a  maximum
redshift of $z< 3.2$ at 90$\%$ confidence level.

Moreover, by using the temporal decay
value of $\alpha=0.4$ obtained by the simple power law fit discussed in
\S~\ref{sec:opticalIR}, the minimization
returns $\beta_{O} = 1.42$.
The results obtained by using this value for $\alpha$
are shown in Figure \ref{fig:redshift}. Here the top panel shows 
the
derived AB magnitudes, corresponding to $BRi'K$ values, and the
best-fitting
curve; the bottom panel  shows the reduced $\chi^2$ distribution
as a function of photometric redshift. The $\chi^2$ distribution does not
show a minimum: it is very
flat from $z\sim0$ to $z\sim2$ and then start increasing if $z>2.4$.
Therefore, it does only allow setting an upper limit determination for 
$z$.

The numerical technique used here does not account for intrinsic 
absorption.
In the attempt to estimate this effect, we imposed a power-law on the
whole SED
from the optical to X-ray wavelength range, and used a SMC extinction
profile. By taking $z\sim2$, we find $A_V = 0.26^{+0.09}_{-0.11}$ 
and $\beta_{OX}
= 0.54^{+0.23}_{-0.12}$. The shallower value of $\beta_{OX}
= 0.54^{+0.23}_{-0.12}$ is
preferable:  if $\beta_{O}$ is as steep as $\sim 1.1$ we would be unable
to account for the SED from optical to X-ray according to the synchrotron
spectrum of the afterglow. Moreover, $\beta_{OX}\,=\,0.54^{+0.23}_{-0.12}$ 
is 
fully
consistent with the X-ray slope registered at the same time, as shown in
Table~\ref{tab:xindices}.

\section{Discussion}
\label{sec:discussion}

\subsection{Afterglow Model}
\label{ag}

The X-ray afterglow of GRB~060108 is quite complex and exhibits
several of the interesting features which have been recently
observed in others GRBs by \textit{Swift} (see \citealt{zha06},
\citealt{no05}, \citealt{bur05b}, \citealt{rom05} for detailed
discussions). In particular, in several cases the X-ray emission has
been found to consist of an initial steep decay, followed by a
shallow one about 100-1000~s after the trigger. A second change of
the decay slope usually occurs about 10 ks later. Flares can be
observed superimposed to this template behavior, and they are
usually thought to be produced by internal shocks that should occur
during the prompt emission phase.

In the case of GRB~060108 we have registered three different
spectral slopes. The X-ray lightcurve initially follows the standard
pattern of rapid to shallow transition, and then, after $t_{\rm
break,2}\sim 11.5$~ks, the afterglow enters the ``normal afterglow''
phase. Also, a small X-ray flare has been observed at $t\sim 300$~s.  

According to the standard interpretation, the initial fast decay is
likely to be associated to the tail of the prompt emission. In such
a case, the relationship between the spectral and decay indices
should be given by $\alpha =2 + \beta$ (``curvature relation'', see
\citealt{kp00}, \citealt{zha06}). In the case of GRB~060108 it is
$\alpha_1\,=\,2.78\,\pm\,0.43$ and $\beta_1\,=\,1.47\,\pm\,0.42$,
therefore the agreement is only marginal.
The value of $\alpha $ required for the curvature relation to be
satisfied is $\sim$\,3.5, although the 
shallower initial slope observed in our case may possibly be caused by the presence 
of the flare at $\simeq$\,300\,s. In fact, the $\alpha =2 + \beta$ 
relation gives the steepest decay index for a spherical model
and is expected to be satisfied only when 
emission from on-axis fluid elements is shut off suddenly, and off-axis 
emission dominates. The presence of minor internal shocks at 
later times would make the decay more gentle. On the other hand, Figure
\ref{fig:batxrtcurve} shows that the backward extrapolation of the
XRT data well matches the flux level of the prompt emission, as
registered by BAT and extrapolated in the XRT band, strengthening
the idea that the initial fast decay of the X-ray emission is
related to the prompt emission.

Just after the rapid decay phase, we observe a small flare that peaks at
300\,$\pm$10\,s and, when modelled with a Gaussian, has a width 
$\sigma_G = 32.3 \pm 10.0$~s. By adopting different rebinning for the X-ray lightcurve, 
we have found that the 
$\chi^2$ always shows an improvement by adding a Gaussian to the fitting model, 
and that the confidence level for the flare is $\sim 2-2.5 \sigma$. 
Such X-ray flares are common features of Swift XRT light curves 
(\citealt{zha06}, \citealt{li06}), and 
are usually associated with collisions between ultrarelativistic shells 
emitted by the ``inner engine'' after  the main high-energy event. In this 
case, the decay slope and the spectral index of the flare should
again satisfy the curvature relation, at least provided that the
decay slope is computed by re-setting the time-zero point t$_{0}$ at
the start of the shell emission (see \citealt{zha06} and references
therein). Although the flare statistic is quite low and does not allow a
robust determination of the slope, we find that this relation
is satisfied by taking $t_0 = 282$~s. This gives $\alpha_{fl}= 4.7 \pm
1.1$ (90\% confidence level, while the
spectral index is $\beta=1.7\pm 0.4$, see Table~\ref{tab:xindices}).

After $t_{\rm break,1} =  250 \pm 30$~s, the GRB~060108 afterglow
enters a phase of plateau, which we can attribute to the emission
from the forward shock produced when the fireball runs into the
circumburst medium (see \citealt{zha06}, \citealt{obr06}). However,
the observed temporal decay slope is shallower than that expected
from a standard afterglow decline (\citealt{spn98}), requiring that
additional energy is injected into the afterglow (``refreshed''
afterglow). Three possible mechanisms have been proposed to explain
this late energy injection (\citealt{zha06} and references therein):
i) a continuous activity from a longer lasting central engine,
usually mimicked by a central engine luminosity law of the kind
$L\propto t^{-q}$, ii) a power-law distribution of the Lorentz
factors in the ejecta, causing slower shells to catch up with the
decelerated fireball or iii) the deceleration of highly magnetized
ejecta, in which case the outflow is Poynting flux dominated and
transfers energy to the medium. In the latter case, however, a good
parametrization of the energy release is still lacking, therefore we
will not discuss this possibility in further details (see
\citealt{zha05} and \citealt{zha06} for a discussion).

By using the closure relations listed by Zhang et al. (2006; see their
Table~2), we find that the relation which is satisfied by the
spectral and decay index during this phase ($\alpha \sim 0.3$ and
$\beta \sim 0.7$, see Tables~\ref{tab:tempindices} and
\ref{tab:xindices}) is $\alpha= (q-1)+ (2+q)\beta/2$, giving $q=
0.44$. All other possibilities are excluded. This implies that the
fireball is propagating  into a
constant density medium (such as the interstellar medium, ISM) and
that the X-ray frequency lies between the synchrotron frequency and
the cooling frequency ($\nu_m < \nu < \nu_c$). During this epoch,
the central engine does not turn off, but continues to inject energy
in the outflow  according to the law $L \propto t^{-0.44}$. This
result makes unlikely that the late energy injection is due to
electromagnetic emission from an highly magnetized and rapidly
rotating pulsar (which in principle can be formed after the
explosion of the supermassive star producing the GRB), as this would
normally require a flatter injection rate during the initial spind-down phase 
($q \sim 0$), followed by an asymptotic decay with $q\sim2$ (see 
\citealt{zha01}). Similarly, any mechanism causing a nearly constant input of 
energy (for example early phases of disk accretion, during which the accretion 
rate, $\dot M$, can be nearly 
constant, etc ...) can be ruled out. However, 
continued infall of matter onto a newly
formed black hole may occur with $q<1$, since at late times $\dot M$ is 
expected to fall off like a power law. 

As mentioned before, an alternative and observationally
indistinguishable scenario is one in which the
central engine activity is brief, but the ejecta have a distribution of
Lorentz factors. The fastest shells initiate the forward
shock, decelerate, and are successively caught by the slowest
shells producing internal shocks. The
consequent addition of energy in the blast-wave
mitigates the deceleration and the afterglow decay rate.
Assuming a Lorentz factor distribution $M(>\gamma)
\propto \gamma^{s}$, where $\gamma$ is the shell Lorentz factor, the index
$s$ is related to $q$ by $s=-(10-7q)/(2+q)$, which, in this case, gives
$s= -2.83$.

The X-ray light curve breaks steeply at 11.5\,ks. We can exclude
that this is due to the passage of the cooling frequency through the
X-ray band at this time, since in such a case, when comparing
parameters relative to the epochs before and after the time break,
we would expect to observe a relatively large change in the X-ray
temporal slope of $\Delta\alpha = 0.25$, accompanied by a similar
change in the spectral slope of $\Delta\beta = 0.5$. In the case of
GRB~060108, the observed change in the decay slope is $\Delta\alpha
=0.76 \pm 0.12$, while the spectral change is only $\Delta\beta =
0.03 \pm 0.27$ at 90$\%$ confidence level. This means that this
interpretation is not compatible with the data (less than $\sim 3
\sigma)$.

More likely, the second break at $t_{\rm break,2} = 11.5 \pm 0.4$~ks
occurs when the additional energy injection ceases and the afterglow
enters a ``normal'' decay phase. During this phase, temporal and
spectral indices satisfy the closure relation $\alpha = 3\beta/2$,
which again confirms that the fireball is propagating in an ISM-like
medium. This also requires the X-ray frequency to be below the
cooling frequency during the entire duration of XRT observations.
The corresponding index of the power-law energy distribution of
radiating electrons is $p = 2.39$.

The comparison between the optical and X-ray light curve is also
illuminating.  Interestingly, the $i'$ and $R$ band flux after the
first 800~s have a decay slope of $\sim 0.3-0.4$, similar to
what is observed at the same time in the X-ray band. This suggests
a common origin for the X-ray and optical emissions at these epochs.

On the other hand, detailed observations in the $R$ and $i'$ band
may suggest that the optical decay slope may be steeper between 300
and 800~s. This may indicate that we are detecting a decay from an
early optical ``bump'', i.e. that at such early times we are
observing the fading emission from a faint reverse shock, which
propagates backward into the ejecta while they are refreshed by the
late central engine activity \citep{zkm03}. A possible rebrightening
is observed in the $i'$ band only. The significance is low and a monotonic 
decay is not ruled out. On the other hand, it is worth to point out that a 
similar feature has been detected at larger significance 
in the brighter GRB~060206 (\citealt{monf06c}), in which case the $R$ and $i'$  
lightcurves could not be fitted with 
identical functions. If real, the rebrightening in the $i'$ band may de due to 
a second reverse shock, possibly initiated by the collision of the 300~s of
the X-ray flare into the external medium.

The X-ray light curve breaks steeply at 11.5\,ks, while no apparent
corresponding break is observed in the $i'$ and $R$ bands. A time break
produced by the passage of the cooling frequency through the X-ray
spectrum would certainly account for this, since in such a case the break
would be monochromatic.

On the other hand, based on the spectral and temporal analysis, we
find that this is not likely to be the case and the X-ray break is most
probably  due to the cessation of energy injection. In such a case,
and if the optical and X-ray emission were produced by the same
mechanism, a break 
should be expected to
appear in the optical emission, at $\sim t_{\rm break,2}$. Here we only 
notice that
observations are sparse at this time; there are no data in the $i'$
band measured after 5\,ks and the only $R$-band measurement taken
after 11.5\,ks indicates that the light curve does not break. 
The fit of the optical lightcurve with a single power law suggests that 
an optical break may not be required even in correspondence of 
the first X-ray break. 
It is therefore possible that the GRB~060108 behaviour is similar to that 
of other GRBs, which do not present a break in the optical contemporaneous to 
that in the X-ray (see \citealt{pan06}). 
A satisfactory
explanation for this behaviour is still lacking. Alternatively, if 
the $R$-band observation ($R$=23.5~mag taken at 12.67~days after the
trigger) is attributed to emission from the host galaxy of the GRB,
a break in the optical GRB light curve at $t_{\rm break,2}$ would not have 
been observed.
As mentioned before, a significant host galaxy contamination is expected,
since the source appears to be extended in the late VLT image.

\subsection{Observed Darkness}
\label{dark}

\begin{figure}
\includegraphics[width=80mm]{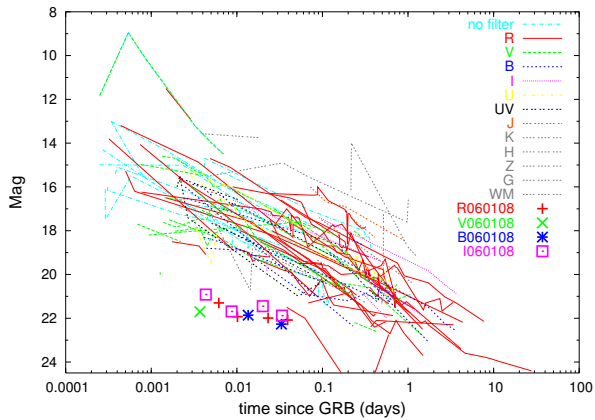}
 \caption{Comparison of GRB afterglow light curves with that
of GRB~060108, illustrating the
observed faintness of its optical emission. Re-adapted from Guidorzi et
al.~(2005); see the original paper for the references of all
data.}
 \label{fig:darkbursts}
\end{figure}

Figure \ref{fig:darkbursts} shows the lightcurve of GRB~060108 in
the broader context of other GRBs. One peculiar characteristic of
this event is that, although the burst is detected in the $BVRi'$
bands at later times, the UVOT does not observe any optical/UV emission
within the first $\sim300$~s (i.e. during the
tail of the prompt emission where internal shocks are expected to
occur).

In order to investigate the possible reason for this behaviour, we
can compare the relative faintness of GRB~060108 with other bursts
using the X-ray flux versus gamma-ray fluence and the optical flux
versus X-ray flux plots published by \cite{rom05}. In the case of
GRB~060108, at 1~hr after the burst the
X-ray flux is $3.1 \times 10^{-12}$erg\,cm$^{-2}$s$^{-1}$
and the gamma-ray fluence is $S=4.9 \times 10^{-7}$erg\,cm$^{-2}$ (in
the 2-10~keV and 15-350~keV bands, respectively).
As we can see
from the Figure 3 of \cite{rom05}, this burst is not unusual and
sits solidly within 2$\sigma$ of the mean, although overall it is
slightly fainter in both $\gamma$ and X-rays compared to other GRBs.
We therefore expect a typical behaviour also in the optical emission
(perhaps with a slightly fainter optical afterglow than other
bursts). If we compare the X-ray and optical fluxes at 1~hour after
the trigger (when the optical flux is $\sim 8.0$~$\mu$Jy),
we find values within the range of the population with a slight X-ray to
optical excess.
Similarly, if we extrapolate the optical point at 11 hours, we find
an optical flux of $\sim 3.9$~$\mu$Jy, while the X-ray flux is $4
\times 10^{-13}$ erg\,cm$^{-2}$s$^{-1}$. Although we caution that the 
behaviour of the light curve at 11~hr is not well constrained (see below), 
if taken face value this would imply  that GRB~060108
still lies in the 2$\sigma$ limit, near the region of optically faint
events.

\begin{figure}
\includegraphics[width=84mm]{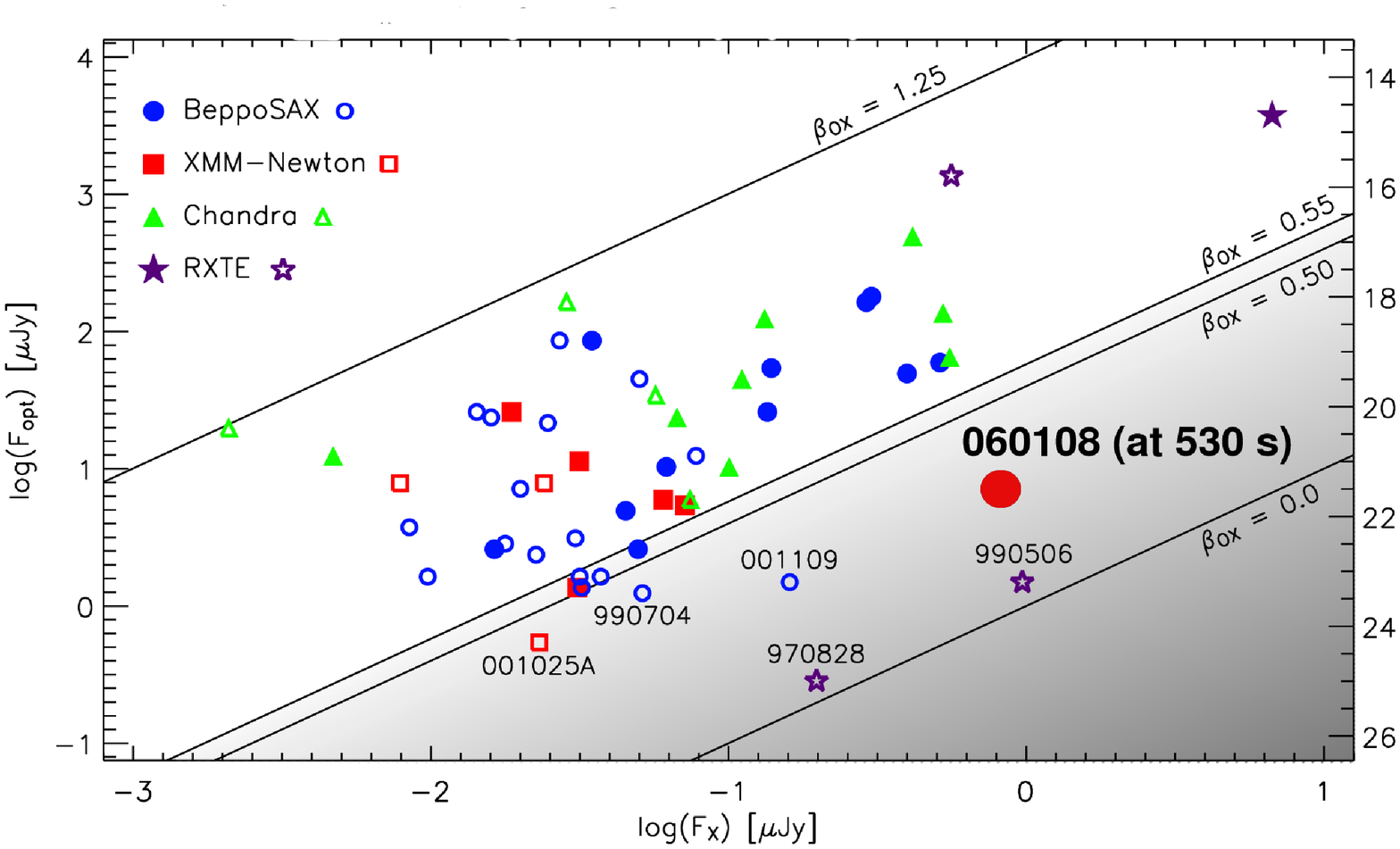}
 \caption{ Ratio of optical to X-ray brightness as a function of time of
GRB~060108, compared with other GRBs, dark and luminous. Re-adapted from
Jakobsson et al. (2004).}
\label{fig:optXcomp}
\end{figure}

A further way to assess the ``darkness'' of a GRB has been
proposed
by \cite{jak04} (see also \citealt{rol05}), on the basis of the optical 
to X-ray spectral
properties. According to these authors, a burst can be classified as
``truly'' dark when $\beta_{OX} < 0.5$. A ratio less than this
threshold would be in disagreement with the spectral slopes expected
from the synchrotron emission, requiring an intrinsic  ``lack'' of
optical emission. Indeed, the ratio of X-ray and $R$ band measurements
of GRB~060108 provide a value of $\beta_{OX}= 0.45$, as measured at
$\sim$\,1000\,s after the burst trigger. According to the
\cite{jak04} classification, this burst is therefore ``dark'' (see
Fig.~\ref{fig:optXcomp}).

In principle, a possible explanation for the ``observed darkness'' is a
large redshift ($z>5$,
\citealt{lr00}, \citealt{kaw06}, \citealt{gen06}). In
such a case the
UV light, which is heavily attenuated by intergalactic hydrogen, is
redshifted into the optical band. The burst would then appear very
red. However, the high redshift scenario is ruled out in this case
as we observe the afterglow in the $B$ band, and we have been able to
determine an upper limit of the redshift of $z<3.2$ by using the
photometric redshift technique. High magnetic fields in the ejecta
may also restrict the amount of energy transferred to electrons by
reverse shocks, inhibiting the optical emission (\citealt{zha05}).

Alternatively, the burst could be obscured by dust. Dust would cause
extinction of the afterglow and the emission appears reddened as
the shorter wavelengths are absorbed more. It is interesting to compare 
the
photoelectric absorption measured from the X-ray spectrum with the
reddening inferred from the optical. As discussed in
\S\ref{sec:resultsgx}, by using all
data after the flare through to the end of the
observations. we have been able to measure
a significant column density excess (99\% confidence level). Assuming
$z=2$, this yields $N_H = 8.6^{+7.2}_{-5.9}\times 10^{21}$~cm$^{-2}$.

Light detected in the $R$ band, centered on 700\,nm in the Earth
restframe, is
emitted at 233~nm in the burst local frame (by assuming again $z \sim 2$).
Assuming a SMC extinction curve, and
$N_{H}/A_{V}=1.6\times~10^{22}$~cm$^{-2}$/mag (\citealt{Weidra00}), the
extinction determined
from the optical-IR SED in Section 3 ($A_V = 0.26^{+0.09}_{-0.11}$) 
corresponds to
an X-ray derived photoelectric absorbing column density of 
$N_{H}=(4^{+1}_{-2})\times
10^{21}$~cm$^{-2}$, consistent with the X-ray measurement. Note however,
that for the SMC metallicity, considerably lower X-ray absorption would be
expected for this $A_{V}$.
This amount of extinction
is
sufficient to bring the optical emission  up to a
value compatible with regular {\it Swift} bursts. Dust extinction
would also alter the spectral index of the optical decay. Our estimated
value, if accounted for, would
bring the Jakobsson flux ratio above 0.5, thus removing this
burst from the optically dark regime.

The possibility that the ``observed optical darkness'' is 
caused by intervening
dust along the line of sight is also favoured by results from the
optical - IR SED modeling. As discussed in \S~\ref{redshift},
without accounting for extinction the technique gives a steep
$\beta_O = 1.42 $, that would be hard to explain in terms of a
synchrotron spectrum of the afterglow. When  extinction is included,
the value of $\beta_{OX}$ is $0.54^{+0.23}_{-0.12}$, which compares
better with the observed SED. It can be noticed that the
value of $\beta_{OX}$ is already low, close to synchrotron emission
limit of $0.5$, even after correcting for extinction. The hardness of the 
spectrum is another factor that contributes to make this burst optically 
faint. 

It would then appear from the arguments above that a likely explanation
for the observed darkness is a combination of intrinsic faintness, 
hard optical to X-ray spectrum  and dust extinction.

\section{Conclusions}

We presented and discussed the gamma-ray, X-ray and
UV/optical/infrared observations of the \textit{Swift} GRB~060108,
performed with the instruments on-board the spacecraft as well as
the with the ground-based Faulkes Telescope North, United Kingdom
Infrared Telescope and Very Large Telescope. GRB~060108
 has a moderately faint X-ray afterglow, but very low optical
emission, making it one of the ``darkest'' GRBs ever observed. We
suggest that faintness of the optical emission may be due to  an intrinsic
weakness of the burst and a hard optical to X-ray spectrum, accompanied 
by 
some degree of
extinction which occurred in the GRB surroundings.

 The X-ray light curve shows the typical template discovered by
{\it Swift}, characterized by a rapid decay in the first 300~s (commonly
interpreted as the tail of the prompt emission on the basis of the
combined temporal/spectral properties) followed by a flat decay slope
(typical of ``refreshed'' afterglows).
An X-ray flare is observed at $\sim300$~s, although the statistic is low. 
The flat decay phase lasts for about $\sim$10~ks, after that
the light curve breaks into a steeper segment with a power-law decay
index of $1.0$, typical of a standard afterglow phase.\\

 A comparison with the optical light curve is quite interesting. While
the optical flux after 800 seconds has a slope of $\alpha\,=\,0.4$,
similar to that of the X-ray in the same interval, the decay index
before this time is likely to be steeper. This behavior may be
explained if we interpret the fast decay as the reverse shock
emission, while the later, flatter emission, has the same origin of
the X-ray one. A possible rebrightening is visible only in the infrared, 
although a monotonic decay
is not ruled out. A similar feature has been seen in the case of GRB~060206 
(\citealt{monf06c}). If real, it might indicate a second reverse shock 
emission, initiated by the
300~s flare. Alternative models able to explain an optical
rebrightening include an increase in the density of the medium where
the forward shock is produced (\citealt{laz02}), or energy injection
by late shells (\citealt{bjo04}, \citealt{joh06}). This last
scenario is less likely, because of our finding that the late energy
injection is taking place at a steady rate. The coincidence of the
rebrightening with the X-ray flare, however, favours the hypothesis
of reverse shock.

Another intriguing feature of the optical light curve is the
absence of a break in correspondence to the X-ray one at $\sim$10\,ks,
which may be either intrinsic or due to a significant contribution to the
optical flux at late times by the host galaxy. However, the
poor sampling of the late optical light curve
does not allow us to better constrain this behaviour. 

 The analysis of the optical spectrum, obtained from data gathered
45 minutes after the burst, has allowed us to determine an upper
limit of $z < 3.2$ at 90\% confidence level, by using a $\chi^2$
minimization of the observed spectral energy distribution.

The optical afterglow is below the detection limit of the UVOT
within 100\,s of the burst. 
This event has shown how observations taken
promptly and deeply enough may reveal the interesting behaviour of the 
early optical emission. 
In the \textit{Swift} era, further similar
observations are a reality due to the prompt response of both the
spacecraft and ground robotic telescopes, as well as due to the
possibility to perform deep observations with large telescopes at
early times after the trigger.

\section{acknowledgements}
CGM acknowledges financial support from the Royal Society.
SZ thanks PPARC for support through an Advanced
Fellowship. KP thanks PPARC for support. AM
acknowledges financial support from the Provincia Autonoma di Trento. The 
Faulkes Telescope is operated 
with support from the Dill Faulkes Educational Trust. The 
UKIRT is operated by the Joint Astronomy Centre on behalf of the UK PPARC. 
VLT observations were carried out under programs 076.A-0392
and 076.D-0667. We acknowledge excellent support from the observing
staff in Paranal. We thank the referee, Prof Ralph Wijers, for useful comments 
and suggestions.

\end{document}